\documentclass[twocolumn, prl,superscriptaddress]{revtex4-1}
\usepackage{xcolor}
\definecolor{red}{rgb}{0.75,0,0}
\definecolor{blue}{rgb}{0,0,0.75}
\definecolor{green}{rgb}{0,0.5,0}
\usepackage{amsmath}
\usepackage{amssymb}
\usepackage{bm}
\usepackage{graphicx}
\usepackage{verbatim}
\usepackage{hyperref}
\usepackage{changes}
\usepackage{lipsum}

\def\be{\begin{equation}}
\def\ee{\end{equation}}
\def\bea{\begin{eqnarray}}
\def\eea{\end{eqnarray}}

\def\besub{\begin{subequations}}
\def\eesub{\end{subequations}}

\def\bwd{\begin{widetext}}
\def\ewd{\end{widetext}}

\definecolor{MediumBlue}{RGB}{83,148,184}

\newcommand{\bsf}[1]{\textsf{\textbf{#1}}}

\definecolor{MediumBlue}{RGB}{83,148,184}

\newcommand{\AM}[1]{\textcolor{black}{#1}}

\newcommand{\AMR}[1]{\textcolor{black}{#1}}

\newcommand{\npgsection}[1]{

\vspace{2ex}\noindent\textbf{\sffamily #1}

\vspace{.5ex}
}
\newcommand{\npgsubsection}[1]{

\vspace{.5ex}\noindent\textbf{\sffamily #1}

}

\title{Two-dimensional chiral active fluids}

\begin{document}
\title{\AMR{Spontaneous rotation can stabilise ordered chiral active fluids}}
\author{Ananyo Maitra}
\email{nyomaitra07@gmail.com}
\affiliation{LPTMS, CNRS, Univ. Paris-Sud, Universit\'e Paris-Saclay, 91405 Orsay, France}
\author{Martin Lenz}
\email{martin.lenz@u-psud.fr}
\affiliation{LPTMS, CNRS, Univ. Paris-Sud, Universit\'e Paris-Saclay, 91405 Orsay, France}
\affiliation{MultiScale  Material  Science  for  Energy  and  Environment,
UMI  3466,  CNRS-MIT,
77  Massachusetts  Avenue,
Cambridge,  Massachusetts  02139,
USA}

\begin{abstract}
Active hydrodynamic theories are a powerful tool to study the emergent ordered phases of internally driven particles such as bird flocks, bacterial suspension and their artificial analogues. While theories of orientationally ordered phases are by now well established, the effect of chirality on these phases is much less studied. In this paper, we present the first complete dynamical theory of orientationally ordered chiral particles in two-dimensional incompressible systems. We show that phase-coherent states of rotating chiral particles are remarkably stable in both momentum-conserved and non-conserved systems in contrast to their non-rotating counterparts. Furthermore, defect separation---which drives chaotic flows in non-rotating active fluids---is suppressed by intrinsic rotation of chiral active particles. We thus establish chirality as a source of dramatic stabilization in active systems, which could be key in interpreting the collective behaviors of some biological tissues, cytoskeletal systems and collections of bacteria.
\end{abstract}
\maketitle

Biological systems are driven out of equilibrium by a continuous supply of energy at the scale of constituent particles. This nonequilibrium driving generates macroscopic forces and currents that are responsible for diverse phenomena ranging from rotation of the cell nucleus \cite{nuc_rot} to motion of tissues to flocking of starlings \cite{Cavagna}. Active hydrodynamics \cite{RMP, Sriram_rev, JSTAT, Prost_Nat}, which augments traditional theories of ordered fluids with extra ``active" terms arising from nonequilibrium driving, presents a general continuum framework to model the macroscopic behaviour of such microscopically driven {systems}.

Orientationally ordered phases in such active systems have properties that are qualitatively distinct from their equilibrium counterparts. This includes the existence of long-range order in active versions of two-dimensional X-Y models \cite{Toner_Tu}, anomalously large number fluctuations \cite{RMP, Aditi2}, and a generic instability whereby an ordered phase of active particles devolves to a spontaneously flowing state in active momentum-conserved fluids \cite{Aditi1, Voit, nuc_rot}. 

Despite their variety, previous studies of active orientable fluids are overwhelmingly restricted to achiral assemblies of particles. Biological objects are however generally chiral, and this microscopic chirality is manifested at all scales up to the cellular \cite{Bershadski} or even at multicellular \cite{cell-layer PNAS, Silberzan} levels. \AMR{Further, organisation of sperm cell into rotating vortices \cite{spKruse, Yang} and circular motion of bacteria at a planar surface \cite{bac1} have been reported.} While some authors have considered \AMR{``circle swimmers" in the absence of fluid hydrodynamics \cite{Yang, Denk, chi_TT, chi_rev, chi_eff, circ1, circ3} and} three-dimensional chiral active fluids~\cite{seb3, Strempel, Cates_drop}, the dynamics of many two-dimensional biological or \AMR{biomimetic} chiral systems \AM{in the presence of hydrodynamic interactions ranging} from epithelia to confined bacterial suspensions remain poorly understood.

In this paper, we consider orientationally ordered phases of such systems in a wide array of experimentally realisable dynamical settings. We consider both nematic and polar objects in two-dimensional momentum-conserved fluids, two-dimensional systems at the interface between two fluids, and suspensions in contact with substrates acting as momentum sinks. We study systems that break top-bottom inversion symmetry, which allows for new, strictly two-dimensional chiral stresses and active rotation of individual particles. Such systems may display ordered phases where many particles rotate in phase in the plane, and we show that a fast enough rotation suppresses the generic instability in these systems, allowing long-range order under most conditions, including momentum-conserved systems where it was hitherto assumed that all orientationally ordered phases are generically destroyed by active forcing. We also demonstrate that another destabilizing mechanism, the activity-driven separation of $\pm 1/2$ defects in systems with nematic symmetry, is also suppressed by autonomous particle rotation. Indeed, the self-propelled $+1/2$ defect generically moves in circles, which prevents it from separating ballistically from a $-1/2$ defect. In chiral, yet non-rotating systems, the $+1/2$ defects do propagate ballistically, albeit at an angle relative to the direction of their polarity. In the following, \AMR{we} present a detailed study of the prototypical case of apolar chiral particles in a momentum-conserved two-dimensional film, and then discuss how changing the modalities of momentum exchange with the substrate and/or introducing polar particles modify our results.
\AMR{\npgsection{Results}
\npgsubsection{Dynamical equations for an apolar chiral fluid}}
\noindent We describe our apolar suspension by a nematic order parameter that depends on the two-dimensional position vector ${\bf r}$:
\begin{equation}
{\bsf Q}({\bf r}, t)=\frac{S}{2}\begin{pmatrix}\cos2\theta & \sin2\theta\\\sin2\theta & -\cos2\theta\end{pmatrix},
\end{equation} 
where $S$ denotes the degree of orientation and $\theta({\bf r}, t)$ is the orientation angle of the apolar particles with respect to the ${x}$ axis. In the absence of fluid flow and activity, the relaxational dynamics of the system is governed by the standard Landau-de Gennes free energy functional, which we write in a single Frank constant approximation for simplicity:
\begin{equation}\label{eq:Hamiltonian}
\mathcal{H}=\int d^2{\bf r}\left[{\alpha\over 2}{\bsf Q}:{\bsf Q}+{\beta\over 4}({\bsf Q}:{\bsf Q})^2+{K\over 2}(\nabla{\bsf Q})^2\right].
\end{equation}
This free energy favours an orientationally ordered phase when $\alpha<0$. We further introduce the hydrodynamic velocity field ${\bf v}({\bf r}, t)$ and its symmetrized gradient tensor $A_{ij}=\partial_iv_j+\partial_jv_i$. Including all terms allowed by symmetry to leading order in gradients, the evolution equation for the order parameter generically reads
\begin{equation}
\label{apol}
D_t{\bsf Q}=-\Gamma_Q{\bsf H}+\lambda{\bsf A}-\lambda_c\boldsymbol{\epsilon}\cdot{\bsf A}-2\Omega\boldsymbol{\epsilon}\cdot{\bsf Q}
\end{equation}
where $D_t$ denotes the co-rotational derivative and $\boldsymbol{\epsilon}$ is the two-dimensional Levi-Civita tensor. The first term on the right-hand-side of Eq.~\eqref{apol} describes the passive relaxation of ${\bsf Q}$ under the influence of the passive torque ${\bsf H}=\delta \mathcal{H}/\delta{\bsf Q}$ deriving from Eq.~\eqref{eq:Hamiltonian}, and the second term expresses the tendency of the apolar particles to orient along the velocity gradient. The next two terms are explicitly chiral: the third is a coupling between orientation and flow allowed in passive fluids, while the fourth is specifically active and 
describes the intrinsic rotation of chiral particles. In a perfectly ordered phase without any flow or concentration gradient, this last term makes the ordering direction rotate globally at a constant rate: $\partial_t\theta=\Omega$.

\begin{figure}
 \includegraphics[width=\columnwidth]{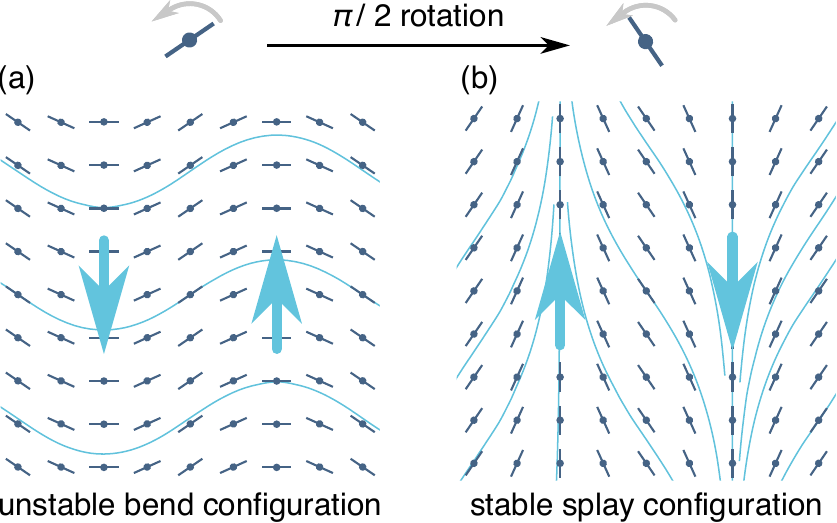}
\caption{Stabilisation of the ordered phase due to the autonomous rotation of chiral particles. (a)~The achiral force $\zeta\nabla\cdot{\bsf Q}$ (blue arrows) destabilizes a bent configuration by inducing a torque that tends to enhance the distortion relative to a perfectly horizontally aligned state. (b)~As the particles rotate by an angle $\pi/2$, however, the bent configuration is converted to a splayed configuration. This results in a reversal of the achiral force, which now suppresses the distortion. Depending on parameters, the superposition of these opposite trends may result in a overall stable system.
}
\label{fig1}
\end{figure}

As the order parameter deviates from this perfectly ordered configuration, flows are generated according to the force balance equation
\begin{equation}
\label{vel}
-\eta\nabla^2{\bf v}=-\nabla\Pi+\zeta\nabla\cdot{\bsf Q}-\zeta_c\nabla\cdot(\boldsymbol{\epsilon}\cdot{\bsf Q}),
\end{equation}
where $\eta$ is the viscosity, $\Pi$ the pressure enforcing the incompressibility constraint $\nabla\cdot{\bf v}=0$, and $\zeta$ and $\zeta_c$ are respectively achiral and chiral active coefficients. The achiral active force stems from the fact that a local polarity leads to a local force in active systems; since $\nabla\cdot{\bsf Q}$ implies a local polarity (associated to bend or splay in two dimensions), it leads to a local force along or opposite this polarity depending on the sign of $\zeta$ \cite{Aditi1}. By contrast, the $\zeta_c$ term characterizes the fact that in the absence of a left-right symmetry, a local polar distortion can lead to a force in the direction \emph{transverse} to it. Equation~\eqref{vel} ignores passive force densities, which are proportional to $\nabla\cdot{\bsf H}$ due to Onsager symmetry, as they are higher order in gradients than the active force densities. It also uses a simplified form for the viscous dissipation, taking the viscosity $\eta$ to be a scalar instead of a more general ${\bsf Q}$-dependent rank four tensor. This simplification does not change any of our results qualitatively even if an antisymmetric ``odd viscosity'' specific to chiral nonequilibrium systems is included~\cite{avron, Debarghya}, as this term ends up being a pure gradient and is therefore irrelevant to the dynamics of incompressible flows.
\AMR{\npgsubsection{Enhancement of linear stability by chiral rotation}}
\noindent Having set up the dynamical equations, we investigate the stability of a homogeneous ordered phase with a constant $S=1$ rotating at a rate $\Omega$. We thus write $\theta({\bf r}, t)=\Omega t+\delta\theta({\bf r}, t)$ where $\delta\theta$ denotes a small perturbation. Using Eq.~\eqref{vel} to eliminate the velocity field, Eq.~\eqref{apol} yields an evolution equation for the angle field $\delta\theta_q$ in Fourier space. To zeroth order in wavevector ${\bf q}$, it reads
\begin{multline}
\label{fluc1}
\partial_t\delta\theta_q=
-\frac{1}{2\eta}\{\zeta\cos [2(\phi-\Omega t)]+\zeta_c\sin [2(\phi-\Omega t)]\}\\
\times\{1+{\lambda}\cos [2(\phi-\Omega t)]-{\lambda_c}\sin [2(\phi-\Omega t)]\}\delta\theta_q,
\end{multline}
where $\phi$ is the angle that ${\bf q}$ makes with the $x$ axis, implying that $(\phi-\Omega t)$ is the angle between the wavevector direction and the instantaneous ordering direction. When both $\zeta_c$ and $\Omega$ vanish, Eq.~\eqref{fluc1} leads to the well-known $\mathcal{O}(q^0)$ mode of an active nematic \cite{Aditi1}. This mode has a positive $q$-independent growth rate for some values of $\phi$, irrespective of the value of $\zeta$, reflecting the well-known generic instability of active suspensions to either bend (perturbation along the ordering direction) or splay (perturbation transverse to the ordering direction)~\cite{RMP, Aditi1}. Now considering a non-zero $\zeta_c$ at $\Omega=0$, the ordered phase is still generically unstable as \AMR{the relaxation rate} $(\zeta\cos 2\phi+\zeta_c\sin 2\phi)(1+{\lambda}\cos 2\phi-{\lambda_c}\sin 2\phi)$ is always negative for some value of $\phi$ irrespective of the parameter values. \AMR{This can be seen by noting that the relaxation rate vanishes at least at $\tan2\phi_0=-\zeta/\zeta_c$ and depending on the sign of the constant $\cos2\phi_0[\zeta_c+(\lambda\zeta_c+\zeta\lambda_c)\cos2\phi_0]$, must therefore be negative either for $\phi\lesssim\phi_0$ or for $\phi\gtrsim\phi_0$. If $\lambda^2+\lambda_c^2<1$, this is also the only condition for a vanishing relaxation rate and therefore the $\phi$ angle sector where the fluid is unstable would be exactly equal to the sector in which it is stable. When $\lambda^2+\lambda_c^2>1$ the relaxation rate also vanishes for wavevectors at which $(1+{\lambda}\cos 2\phi-{\lambda_c}\sin 2\phi)=0$ thereby making the $\phi$ angle sector where the system is stable larger (or smaller) than the sector where it is unstable. The instability of chiral fluids is qualitatively similar to the generic instability of achiral fluids with one crucial difference: unlike in active nematics, its growth rate is not invariant under $\phi\to -\phi$, reflecting the breaking of the left-right symmetry by chirality. As a result of this symmetry breaking, when chiral fluids are confined to and strongly anchored at the boundary of a straight \cite{Voit, Silberzan} or annular \cite{nuc_rot, Sebastian_rot, Bershadski} channel whose width is larger than the ($\propto 1/\sqrt{\zeta}$) length-scale for active instability, left-tilted and right-tilted patterns are no longer equally probable.}
Now considering $\Omega\neq 0$ systems, we find the rotation of the particles can suppress this generic instability. This stabilization is most easily understood in the fast-rotating limit where $\Omega\gg (\zeta S)/2\eta$ and $\Omega\gg(\zeta_c S)/2\eta$, meaning that the angular frequency of the particles is much larger than the growth rate of the generic instability. Averaging over this fast rotation, we obtain the effective long-time dynamics of the angular fluctuations as
\begin{equation}
\label{avfluc}
\partial_t\delta\theta_q=-\frac{\zeta\lambda-\zeta_c\lambda_c}{4\eta}\delta\theta_q.
\end{equation}
In this limit, angular fluctuations thus have an isotropic relaxation rate independent of the wavevector magnitude. That rate is positive if $\zeta\lambda>\zeta_c\lambda_c$, leading to the stabilisation of the ordered phase. Since $\delta\theta$ is also the phase of the angular rotation of the particles, this implies that the fluctuations about the phase-coherent state is damped out by the fluid-mediated interactions \AMR{(see section II A of \cite{supp} for a more detailed description of this stabilisation)}. To understand this stabilization, consider a system that, for $\Omega=0$, is unstable to bend ($\phi\simeq 0$) deformations and stable to splay ($\phi\simeq\pi/2$). In such a system, when the global ordering direction rotates at the rate $\Omega$, a bend deformation becomes a splay deformation after the rotation of the global ordering direction through $\pi/2$ (Fig.~\ref{fig1}). Thus, for fast enough rotation, the unstable configuration is converted into a stable one before the instability has had time to grow. In general, if for $\Omega=0$ the \AMR{the integral of the growth rate over all $\phi$, \emph{i.e.}, all possible wavevector directions, is negative}, rotation of the particles at a fast enough $\Omega$ stabilises it. \AMR{This is reminiscent of the stability of a ball on a rotating saddle. A similar stabilisation mechanism is also used in Paul traps or quadrupolar ion traps; since static electric fields are solenoidal and have both stable and unstable directions these devices trap ions by a fast switching of the stable and the unstable directions.} Further, since Eq.~\eqref{avfluc} is independent of $|{\bf q}|$, the relaxation rate of angular fluctuations does not vanish even for infinite systems when $\zeta\lambda>\zeta_c\lambda_c$. As a result, when modelling a noisy system by adding a small noise to \eqref{apol} we find that the root mean square value of $\delta\theta$ remains finite for $q\to 0$ in two dimensions \AMR{\emph{i.e.}, $\lim_{{\bf r}\to \infty}\langle [\theta({\bf r},t)-\theta(0,t)]^2\rangle$ does not diverge}. This implies that the system has long-range order even in two dimensions. This chirality-induced spontaneously rotating phase is the only known long-range ordered state in two-dimensional momentum-conserved active systems.

\begin{figure}
 \includegraphics[width=\columnwidth]{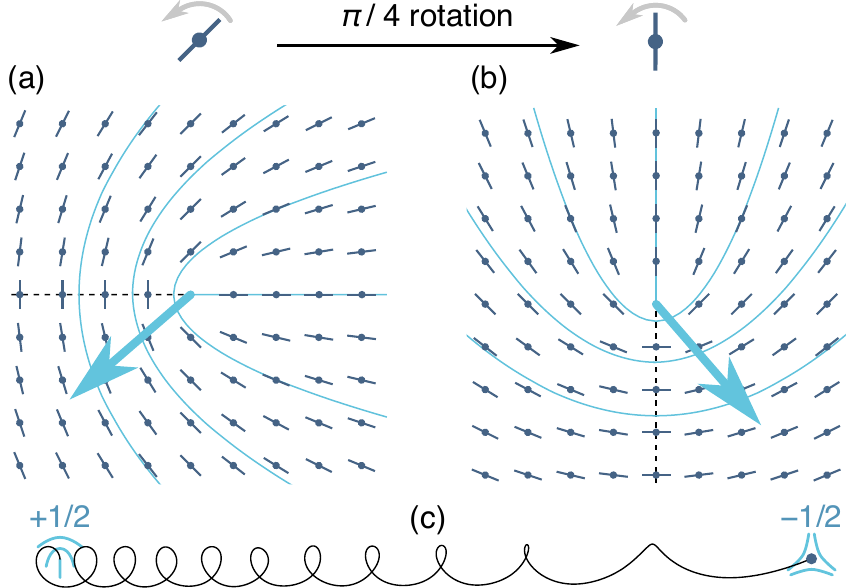}
\caption{\AMR{Rotation of a $+1/2$ defect and consequent recombination with its $-1/2$ partner}. (a)~The combined influence of the achiral ($\zeta$) and chiral ($\zeta_c$) active forces endows a $+1/2$ defect with a self-propulsion velocity (blue arrow) that makes an angle $\tan^{-1}(\zeta_c/\zeta)$ with the main direction of the defect (dashed line). (b)~As each particle rotates autonomously, this main direction rotates as well. The defect velocity thus also rotates at a constant rate $\Omega$, and the self-propelling defect moves in a circle. \AMR{(c)~When in the vicinity of a $-1/2$ defect (dark blue circle, assumed stationary here), the circling $+1/2$ defect is additionally attracted to it, leading to a recombination trajectory materialized by the solid black line (see the supplement~\cite{supp} for the defect equations of motion).}}
\label{fig2}
\end{figure}

\AMR{\npgsubsection{Suppression of defect separation}}
\noindent Chirality also has important consequences beyond linear stability. At high activity (large $\zeta$), the dynamics of achiral active nematics is known to be governed by the proliferation and motion of defects, which induce chaotic flows beyond those predicted by a linear stability analysis \cite{Mishra}. In nematics, the most abundant defects are those with charge $\pm 1/2$, which can be parametrised by
\begin{equation}
\label{param}
\theta=\Omega t\pm\frac{1}{2}\psi,
\end{equation}
where $\psi$ is the angle of the two-dimensional polar coordinates. \AMR{Defect pairs with no net topological charge are spontaneously nucleated even in equilibrium at any finite noise-strength. In equilibrium, all such defects recombine due to the attractive force which scales as the inverse of the separation between the defects. However, in active systems dynamics is not governed by an energy functional and in particular, any geometrically polar object is inevitably associated with a propulsive force either along or opposite to the polarisation direction. A nematic $+1/2$ defect has a polar structure and therefore, in an achiral but active system, moves either in the direction of or opposite to the polarity \cite{Mishra, Pismen, Vijay1}. In contrast, the $-1/2$ defects have a triaxial symmetry, implying that they can only diffuse. As a result, the relative ballistic motion of the two types of defects leads to a separation of $(+\frac{1}{2}, -\frac{1}{2})$ defect pairs at least at low noise (at intermediate noise strengths, the decorrelation of the direction of motion of the $+1/2$ defect due to its rotational diffusivity suppresses defect-separation resulting in a reentrant nematic phase; see \cite{sshankar}). This activity-driven unbinding of charge-neutral pairs at low noise (or high activity) destroys the orientationally ordered phase via an active analogue of Kosterlitz-Thouless transition \AMR{\cite{sshankar}} and eventually results in spatio-temporal chaos. In a chiral system, the self-propulsion velocity of $+1/2$ defect is however markedly different: by introducing Eq.~\eqref{param} into the right-hand-side of Eq.~\eqref{vel}, we can calculate the active force density associated with a $+1/2$ defect oriented along the $x$ axis {at $t=0$} \AMR{(see section IV of \cite{supp} for the detailed calculation)}:}
\begin{multline}
{\bf f}_{+1/2}=
\frac{1}{2r}\{\zeta[\cos (2\Omega t)\hat{x}+\sin(2\Omega t)\hat{y}]\\
-\zeta_c[\sin(2\Omega t)\hat{x}-\cos(2\Omega t)\hat{y}]\},
\end{multline}
where $r$ is the radial distance from the defect core. Using the Oseen tensor appropriate for two-dimensional momentum conserved flows in a domain of size $R$, we calculate the fluid velocity at the origin (\emph{i.e.}, at the centre of the defect), which in turn advects the defect core~\cite{Mishra}:
\begin{multline}
\label{def_vel}
{\bf v}_{+1/2}=
\frac{R}{4\eta}\{\zeta[\cos (2\Omega t)\hat{x}+\sin(2\Omega t)\hat{y}]\\
-\zeta_c[\sin(2\Omega t)\hat{x}-\cos(2\Omega t)\hat{y}]\}
\end{multline}
This is the self-propulsion velocity of a $+1/2$ defect in a chiral apolar fluid.
For non-rotating systems ($\Omega=0$), we see that unlike in achiral systems, the defect does not move along (or opposite) to its polarity but at an angle $\tan^{-1}(\zeta_c/\zeta)$ to it. 
\AMR{In contrast, when $\Omega\neq 0$ the polar axis of the defect rotates systematically [see Fig.~\ref{fig2}(a-b)]. Therefore, {an isolated} $+1/2$ defect does not move ballistically but in a circle of radius $(R/4\eta)\sqrt{\zeta^2+\zeta_c^2}/2\Omega$. This implies that a $+1/2$ defect nucleated at finite distance from its $-1/2$ partner remains within a finite distance of it, as opposed to continuously moving away from it, simply due to the deterministic rotation of its direction of motion. In addition to this rotation, the passive attractive elastic interaction $\propto -(K/r^2)\mathbf{r}$, where $\mathbf{r}$ is the relative position of the defects, promotes the eventual recombination of the $\pm 1/2$ pair. Figure~\ref{fig2}(c) shows a typical deterministic trajectory for this recombination, and we further discuss its dynamics in Sec.~V of the supplement~\cite{supp}. This implies that} beyond conferring linear stability, chirality-induced active rotation also leads to the suppression of chaos-inducing defect-separation.

\AM{It is instructive to compare this mechanism for the suppression of defect separation with the recently proposed one for achiral active nematics~\cite{sshankar}. There, the rotational diffusivity of self-propelling $+1/2$ defects ultimately suppresses defect separation at low but non-zero noise strength. The same mechanism is operative in chiral but \emph{non-rotating} active nematics which implies that defect unbinding in this case is suppressed for low enough activity and intermediate noise-strengths. In contrast, in spontaneously rotating nematics, we have shown that an additional active mechanism, namely spontaneous rotation of individual particles and hence, a coherent spontaneous rotation of the propulsion direction of a $+1/2$ defect, hinders defect-separation. This implies that the re-entrant disordered phase that is generically present in achiral active nematics \cite{sshankar} is absent in this case -- the spontaneously rotating ordered phase persists even at arbitrarily small noise strengths.}
\AMR{\npgsubsection{Generality of our results}}
\noindent The qualitative conclusions described above for momentum-conserved two-dimensional systems also hold for more experimentally accessible two-dimensional interfaces in contact with a three-dimensional fluid. However, the relaxation rate of angular fluctuations in these systems depend linearly on the wavevector magnitude $|q|$:
\begin{multline}
\label{fluc2}
\partial_t\delta\theta_q=-\frac{|q|}{8\eta}\{\zeta\cos [2(\phi-\Omega t)]+\zeta_c\sin [2(\phi-\Omega t)]\}\\\times\{1+{\lambda}\cos [2(\phi-\Omega t)]-{\lambda_c}\sin [2(\phi-\Omega t)]\}\delta\theta_q,
\end{multline}
where the fluids above and below the interface are both taken to have a viscosity $\eta$ for simplicity. As a result, an $\Omega=0$ system is still unstable as ${\bf q}\to 0$. Nevertheless, rotation stabilises the system and leads to a long-range ordered phase as described above \AMR{(see section II B of \cite{supp} for a detailed discussion of this case)}. Our conclusions about the motion of defects are also still valid.

Our basic conclusion regarding the enhanced stabilisation due to the chiral rotation of particles is also valid for a system that exchanges momentum with a solid substrate, \emph{e.g.}, an epithelial cell layer. This case is however more complex than the previous ones, as the force balance equation Eq.~\eqref{vel} is modified in two crucial ways. First, a frictional force $-\Gamma {\bf v}$ is introduced. Second, a further achiral active force $2\zeta_2 {\bsf Q}\cdot(\nabla\cdot{\bsf Q})$ is allowed \cite{Ano_apol}. In this system, the relaxation rate of angular fluctuations is proportional to $\mathcal{O}(q^2)$ due to the presence of friction. As a result, the root mean square value of $\delta\theta$ in this case diverges logarithmically, which in turn implies that the ordered phase only has quasi-long-range order. 
The conclusions reached earlier about the lack of $\phi\to-\phi$ symmetry of the relaxation rate and the motion of defects when $\Omega=0$ remain valid, however. The fast rotations for large $\Omega\neq 0$ again wash away the anisotropy of the angular relaxation rate, leading to an average growth equation~
\begin{equation}
\partial_t\delta\theta_q=-\frac{q^2}{4\Gamma}(2\zeta_2+\lambda\zeta-\lambda_c\zeta_c)\delta\theta_q-\Gamma_QKq^2\delta\theta_q,
\end{equation}
with similar characteristics to Eq.~\eqref{fluc1} \AMR{(see Sec II C of \cite{supp}}. As before, $+1/2$ defects move in circle. Note that while achiral apolar fluids on a substrate are not generically unstable even at high activities~\cite{Ano_apol}, the spontaneous rotation of the chiral particles enlarges the range of parameter in which the chiral system is stable relative to the achiral one. In particular, unlike in the achiral case, neither $\zeta_2>\zeta_1$ nor $|\lambda|<1$ is essential for stability.

Now considering polar particles instead of apolar ones, our conclusions regarding the stability of the long-ranged ordered rotating phase and generic instability of the non-rotating one remain unchanged for both free-standing films or those exchanging momentum with a three-dimensional fluid. Indeed, in these momentum-conserved systems, polarity does not affect the dynamics at lowest order in gradients \AMR{(see section III A and B of \cite{supp})}. {The case where the particles are in contact with a substrate does however qualitatively differ from its apolar counterpart, as it allows for particles that both self-propel and rotate. These particles thus swim in circles. The absence of momentum conservation additionally allows for a coupling between the polarization of the particles and the velocity of the fluid. In the presence of an incompressible fluid, which was ignored in previous treatments~\cite{chi_TT, chi_rev, chi_eff, circ1, circ3}, these couplings can enhance the stability of the ordered phase coherent state of these circle swimmers, leading to a stable long-range ordered phase with a relaxation rate independent of system size~\cite{Ano_pol} \AMR{and section III C of \cite{supp}}.}

The theory presented here may be extended to study the dynamics of concentration fluctuations as well. We perform this analysis in \AMR{section VI of} the supplement~\cite{supp}, and show that the number fluctuations in almost all rotating ordered phases are normal, \emph{i.e.}, the root mean square number fluctuations  in a region containing $N$ particles on average scale as $\sqrt{(\delta N)^2}\approx\sqrt{N}$, as it should in all equilibrium systems. The number fluctuations in the chiral apolar phase on a substrate are however giant and scale as $\sqrt{(\delta N)^2}\approx N$ both when $\Omega=0$ and $\Omega\neq 0$ \AMR{(see section VI B 3 of \cite{supp} for a more detailed discussion of this case)}.
\AMR{\npgsection{Discussions}}
\noindent Our study has implications for recent experiments suggesting that some epithelial cell layers form chiral, albeit non-rotating phases~\cite{Silberzan}. These cellular aggregates display an asymmetry between left and right-tilted patterns when confined in narrow straight channels, as predicted by our theory. Such experiments should additionally allow for the measurement of the direction of propagation of a $+1/2$ defect, which we predict should be tilted at an angle $\tan^{-1}(\zeta_c/\zeta)$ with respect to their polarity axis. As a result, the measurement of this angle can immediately reveal the strength of chiral active forces relative to the achiral ones. 

Beyond these non-rotating systems, chiral rotation has recently been observed in a nematic quasi-two-dimensional layer of microtubules with kinesin motors\AMR{~\cite{Kim, Sano}}. In \AMR{these} system, microtubules locally rotate in synchrony, with their phase being preserved over distances of milimetres \AMR{i.e. hundreds of times the length of microtubules}. This persistence of order is striking when compared with the rapid onset of spatiotemporal chaos in non-rotating microtubule assays~\cite{Dogic}, and could hint at a stabilization of order by rotation. \AMR{Furthermore, increased density of microtubules or enhanced volume exclusion interaction led to the disruption  of the ordered state in \cite{Sano}, as is to be expected since steric interaction and large densities hinder the formation of locally rotating phases.}
Reproducing these rotating phases at a two-fluid interface in conditions similar to those of Ref.~\cite{Dogic} would yield more insights into their relative stability \AMR{as in this case, the non-rotating ordered phase is gnerically unstable even at infitesmal activity}. \AMR{Further, an examination of the motion of topological defects in these experiments would allow for the verification of the predicted circular motion of $+1/2$ defects.}

Finally, our predictions regarding polar chiral swimmers could also be tested experimentally, for instance using \AMR{sperm cells} or bacterial strains known to swim in circles near a solid surface~\cite{bac1, spKruse, bac2}. A dense collection of such \AMR{sperm cells} or bacteria in a confined channel would be an example of a polar chiral fluid. \AMR{We predict that such systems can sustain a state of long-range synchronised circular motion due to hydrodynamic interactions, somewhat akin to the ``vortex array" phase described in \cite{spKruse} (albeit there, each ``vortex" contains multiple sperm cells).} 

\AMR{While our theory provides a generic explanation for stability of rotating active phases in biological systems such as in \AMR{\cite{Kim, Sano}}, a truly discriminating test for our stability mechanism would require decoupling chiral rotation and activity -- it would involve demonstrating that an achiral but active orientationally ordered state is unstable while the same system, with the same level of activity but with chiral rotation remains stable. At present, this degree of control is unavailable in biological systems. However, artificial chiral anisotropic particles \cite{aspp} endowed with a self-propulsion mechanism \cite{gsr1, gsr2, gsr3} may provide an ideal test-bed for such well-controlled and quantitative tests. Such chiral active swimmers can be engineered, for instance, using chemotactic colloids \cite{gsr2, saha} in which the mobility axis, which depends on the shape of the particle, is not coincident with the chemical axis i.e., the axis along which the catalytic coat is applied. A further apparent difficulty for experimental tests of our theory concerns obtaining the synchronised rotating state itself. Liquid crystals are generally thought to orient via steric interactions which however can not lead to synchronisation of rod-shaped rotating particles. This is, however, not as great a difficulty as it appears at first sight: ordering in both polar and apolar active fluids is known to set in even at low densities at which direct steric interactions between the particles is irrelevant \cite{Ano_pol, Harsh, Brotto, ano_unpub} due to hydrodynamic or other interactions which do not depend on physical contact between the particles. Such interactions are possible in chiral rotating systems as well \cite{ano_unpub} and would generically lead to the kind of ordered states that we have considered here and our detailed predictions regarding stability and defect motion may be quantitatively tested, with the various phenomenological coefficients appearing in our models being determined directly from the design parameters of the active particles. More precisely, a measurement of the static structure factor of $\delta\theta$ fluctuations, when the rotating active fluid is stable, would scale as the inverse of the relaxation rate in \eqref{avfluc} or its equivalent expressions for systems on a substrate or at an interface between two fluids.} Overall, we argue that the study of chiral active systems has the potential to overturn the widespread belief that incoherent flows are inevitable in active systems.

\begin{acknowledgments}
AM acknowledges insightful and illuminating discussions with S. Ramaswamy, P. Silberzan and J-F. Joanny. This work was supported by a Marie Curie Integration Grant PCIG12-GA-2012-334053, ``Investissements d'Avenir'' LabEx PALM (ANR-10-LABX-0039-PALM), ANR grant ANR-15-CE13-0004-03 and ERC Starting Grant 677532 to ML. Our group belongs to the CNRS consortium CellTiss.
\end{acknowledgments}

%
%

\end{document}